\documentclass[journal=acsnanozhu,manuscript=article, layout=twocolumn]{achemso}

\usepackage{graphicx}
\usepackage{times,mathptmx}

\usepackage{upgreek}
\usepackage{amsmath}

\usepackage[version=3]{mhchem} % Formula subscripts using \ce{}

\author{Zhuoqing Li}
\affiliation{Institute for Materials and X-Ray Physics, Hamburg University of Technology, Denickestr.\ 15, 21073 Hamburg, Germany}
\alsoaffiliation{Centre for X-Ray and Nano Science CXNS, Deutsches Elektronen-Synchrotron DESY, Notkestr.\ 85, 22607 Hamburg, Germany}
\author{Aileen R. Raab}
\affiliation{Institut für Organische Chemie, Universität Stuttgart, Pfaffenwaldring 55, 70569 Stuttgart, Germany}

\author{Mohamed A. Kolmangadi}
\affiliation{Bundesanstalt f\"ur Materialforschung und -pr\"ufung (BAM), Unter den Eichen 87, 12205 Berlin, Germany}

\author{Mark Busch}
\affiliation{Institute for Materials and X-Ray Physics, Hamburg University of Technology, Denickestr.\ 15, 21073 Hamburg, Germany}
\alsoaffiliation{Centre for X-Ray and Nano Science CXNS, Deutsches Elektronen-Synchrotron DESY, Notkestr.\ 85, 22607 Hamburg, Germany}
\author{Marco Grunwald}
\affiliation{Institut für Organische Chemie, Universität Stuttgart, Pfaffenwaldring 55, 70569 Stuttgart, Germany}

\author{Felix Demel}
\affiliation{Institut für Organische Chemie, Universität Stuttgart, Pfaffenwaldring 55, 70569 Stuttgart, Germany}

\author{Florian Bertram}
\affiliation{Deutsches Elektronen-Synchrotron DESY, Notkestr.\ 85, 22607 Hamburg, Germany}

\author{Andriy V. Kityk}
\affiliation{Faculty of Electrical Engineering, Czestochowa University of Technology, Al. Armii Krajowej 17, 42-200 Czestochowa, Poland}

\author{Andreas Schönhals}
\affiliation{Bundesanstalt f\"ur Materialforschung und -pr\"ufung (BAM), Unter den Eichen 87, 12205 Berlin, Germany}

\author{Sabine Laschat}
\affiliation{Institut für Organische Chemie, Universität Stuttgart, Pfaffenwaldring 55, 70569 Stuttgart, Germany}

\author{Patrick Huber}
\affiliation{Institute for Materials and X-Ray Physics, Hamburg University of Technology, Denickestr.\ 15, 21073 Hamburg, Germany}
\alsoaffiliation{Centre for X-Ray and Nano Science CXNS, Deutsches Elektronen-Synchrotron DESY, Notkestr.\ 85, 22607 Hamburg, Germany}
\email{patrick.huber@tuhh.de}

\title{How do ionic superdiscs\\ self-assemble in nanopores?}

\abbreviations{ILC, LC, PCry, Iso}

\begin{document}

\begin{tocentry}
  
 \centering
	{\includegraphics[width=\textwidth]{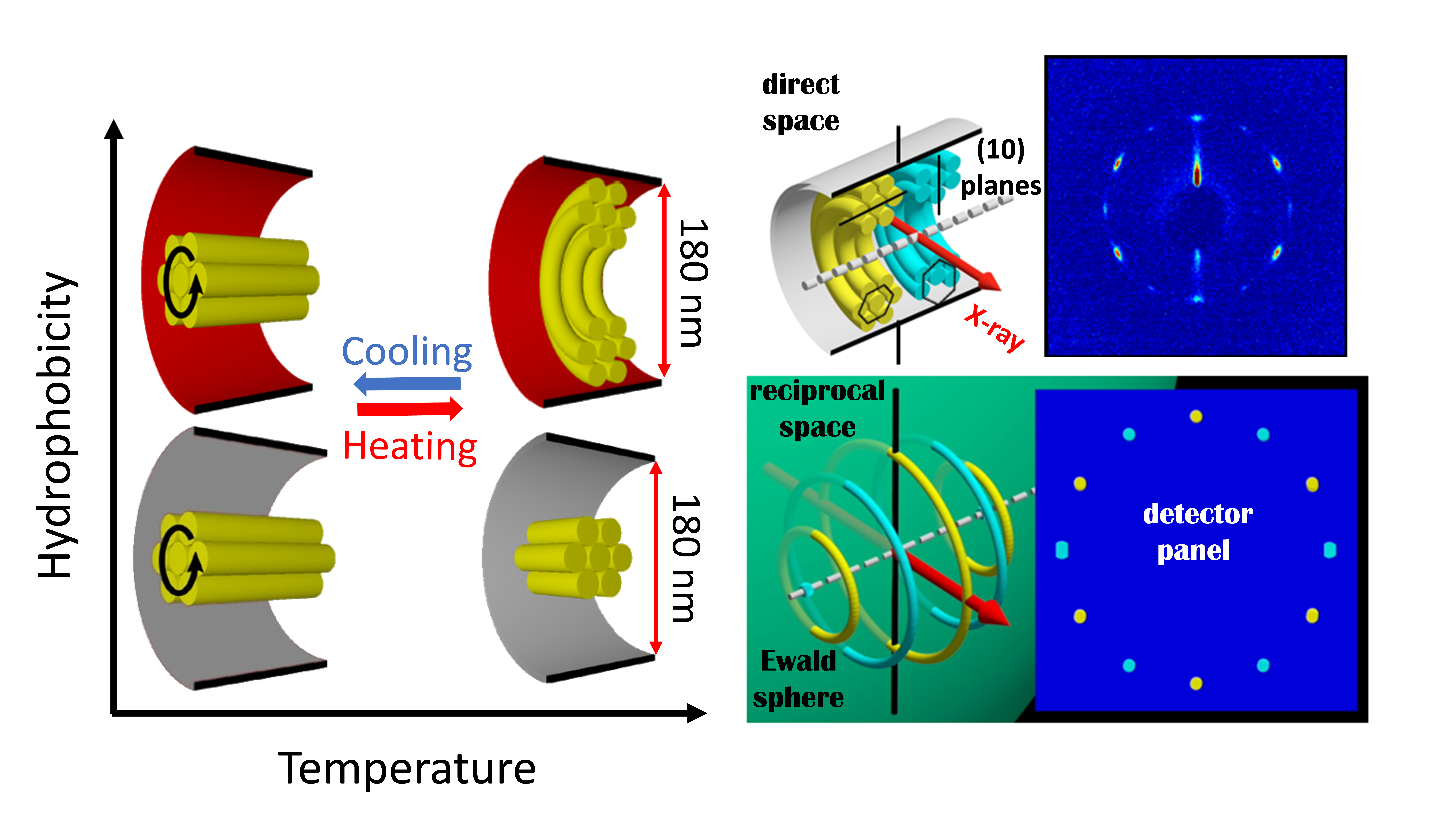}}

\end{tocentry}

\newpage
\begin{abstract}
  Discotic ionic liquid crystals (DILCs) consist of self-assembled superdiscs of cations and anions that spontaneously stack in linear columns with high one-dimensional ionic and electronic charge mobility, making them prominent model systems for functional soft matter. Compared to classical non-ionic discotic liquid crystals (DLCs), many novel liquid crystalline structures with a unique combination of electronic and ionic conductivity have been reported, which are of interest for separation membranes, artificial ion/proton conducting membranes and optoelectronics. Unfortunately, a homogeneous alignment of the DILCs on the macroscale is often not achievable, which significantly limits their applicability. Infiltration into nanoporous solid scaffolds can in principle overcome this drawback. However, due to the extreme experimental challenges to scrutinise liquid crystalline order in extreme spatial confinement, little is known about the structures of DILCs in nanopores. Here, we present temperature-dependent high-resolution optical birefringence measurement and 3D reciprocal space mapping based on synchrotron-based X-ray scattering to investigate the thermotropic phase behaviour of dopamine-based ionic liquid crystals confined in cylindrical channels of 180~nm diameter in macroscopic anodic aluminum oxide (AAO) membranes. As a function of the membranes' hydrophilicity and thus the molecular anchoring to the pore walls (edge-on or face-on) and the variation of the hydrophilic-hydrophobic balance between the aromatic cores and the alkyl side chain motifs of the superdiscs by tailored chemical synthesis, we find a particularly rich phase behaviour, which is not present in the bulk state. It is governed by a complex interplay of liquid crystalline elastic energies (bending and splay deformations), polar interactions and pure geometric confinement, and includes textural transitions between radial and axial alignment of the columns with respect to the long nanochannel axis. Furthermore, confinement-induced continuous order formation is observed in contrast to discontinuous first-order phase transitions, which can be quantitatively described by Landau-de Gennes free energy models for liquid crystalline order transitions in confinement. Our observations suggest that the infiltration of DILCs into nanoporous solids allows tailoring their nanoscale texture and ion channel formation and thus their electrical and optical functionalities over an even wider range than in the bulk state, in a homogeneous manner on the centimeter scale as controlled by the monolithic nanoporous scaffolds.
\end{abstract}

\section{Introduction}
Ionic liquid crystals are salts composed of large organic cations and anions. \cite{Goossens2016, Devaki17} They bridge the gap between conventional ionic liquids and liquid crystals (LCs) and show great potential in fundamental physicochemical research. \cite{Goossens2016, Kapernaum2022} Compared to conventional ionic liquids, ionic liquid crystals (ILCs) often possess long alkyl chains and various functional groups that allow the tailoring of their molecular structure and the corresponding self-assembled mesophases. \cite{Axenov2011} Among the various ionic liquid crystals, those with a discotic structure have attracted particular interest. Discotic ionic liquid crystals (DILCs) often contain aromatic cores in their large cation units. Driven by the $\pi-\pi$ interaction and electrostatic interaction between the aromatic cores, DILCs may self-organise and stack up in columns, resulting in a hexagonal ordered columnar liquid crystalline mesophase, see Fig.~\ref{fig:assem}. \cite{Yildirim2018} The overlapping $\pi$ electrons also provide high one-dimensional charge carrier mobility along the columnar axes often leading to semiconducting properties. \cite{Sentker2018, Bisoyi2022} DILCs therefore combine the high ionic diffusion and conductivity of ionic liquids with the self-assembly and stimulus-responsive anisotropy of conventional discotic liquid crystals. \cite{Kato2017} The unique properties of DILCs allow the design of stimuli-responsive conductors for energy storage devices, electrochromic supercapacitors, flexible batteries, separation media, and optoelectronic devices \cite{Salikolimi2020} In particular, the stimuli-responsive phase transition of DILCs allows the direction of ionic conduction to be modulated. \cite{Devaki17}

\iftrue
\begin{figure}[htbp]
 \centering
	{\includegraphics[width=0.8\columnwidth]{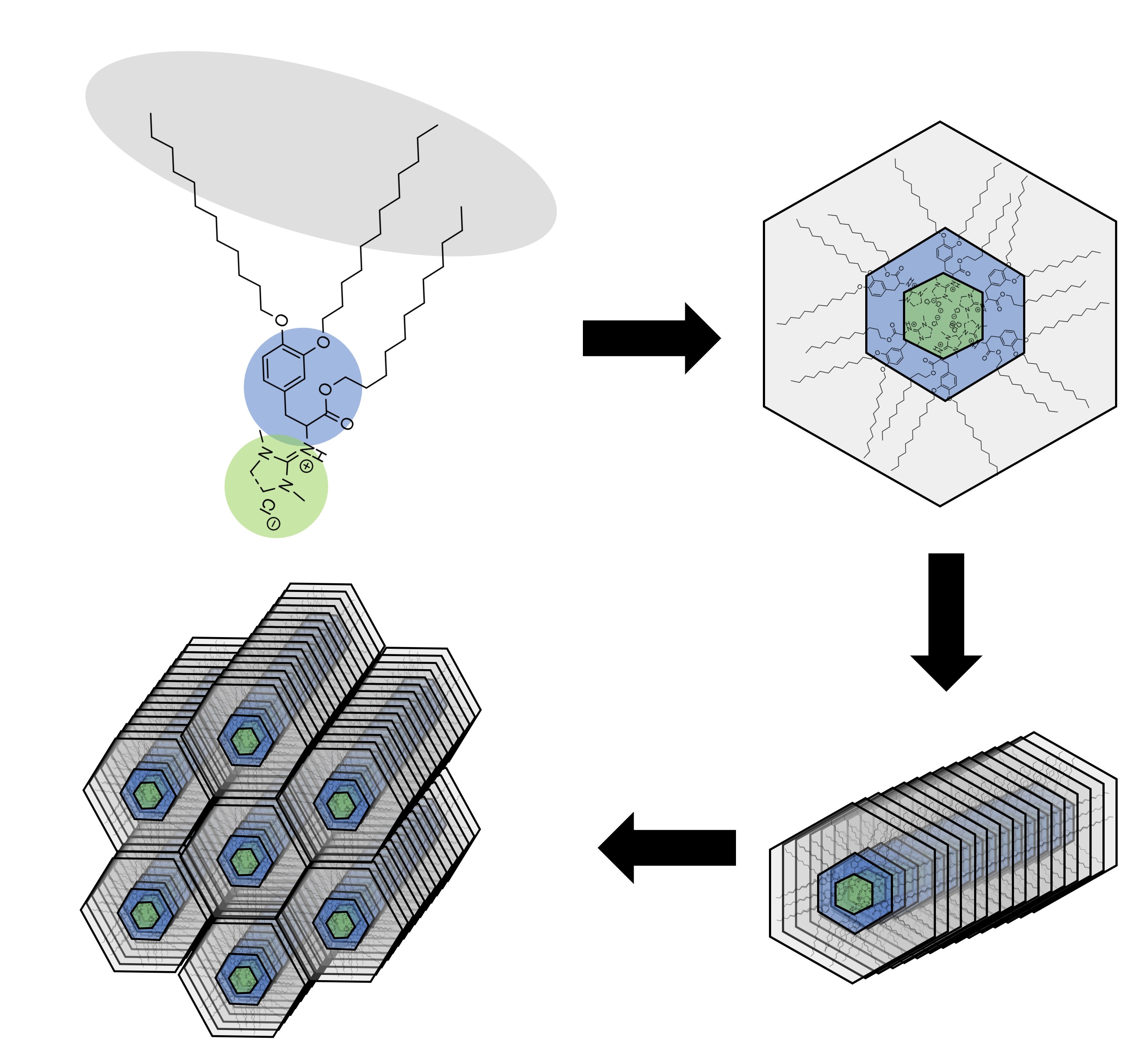}}
	\caption{\textbf{Multiscale self-assembly of a columnar discotic ionic liquid crystal in nanoconfinement.} (a) Chemical formula of a single ionic liquid crystal molecular block based on L-3,4-dihydroxyphenylalanine (Dopa-ILC), consisting of a hydrophilic aromatic core (blue), a chloride anion group (green) and three hydrophobic alkyl chains of different lengths (grey). An acyclic (or cyclic) imidazolium five-membered ring is attached as a functional group to one of the side chains. (b) Six Dopa-ILC molecules self-assemble into a superdisc (discotic unit) with a hydrophilic inner core and surrounding hydrophobic alkyl chains. Calculated from the X-ray diffraction results, the typical size of a superdisc composed of six acyclic Dopa-ILC molecules with a side chain of 12 carbon atoms is 6.85~nm. \cite{C8CP03404D} (c) Due to the $\pi - \pi$ interaction, the discotic units can stack up to form a column. (d) The columns can self-organise further to form a hexagonal columnar liquid crystalline mesophase.}
		\label{fig:assem}
\end{figure}
\fi

Over the past two decades, a large number of synthetic strategies have been established and the properties of DILCs have been studied experimentally and theoretically. \cite{Axenov2011, Woehrle2015, Goossens2016} 
Unfortunately, a homogeneous alignment of the DILCs on a macroscale is often not achievable, which significantly limits their applicability.  A possible reason for this is that electric field-induced orientation, which is the most common orientation method for typical LCs, is not applicable to ILCs due to their ionic nature. \cite{Salikolimi2020} Infiltration into interface-dominated nanoporous solid scaffolds can overcome this shortcoming, as has been demonstrated for many other soft matter systems, most notably non-ionic liquid crystalline materials \cite{Crawford1996, Matthias2005, Kityk2010, Chahine2010, HoRyu2017, Spengler2018, Bisoyi2019, Sentker2019, Sentker2019Dissertation}.

However, because of spatial and topological constraints, the phase and self-assembly behaviour of confined soft matter \cite{Huber1999, Alvine2006Solvent, Alba-Simionesco2006, Schaefer2008, Huber2015} and in particular confined LCs may substantially deviate from the bulk state and DILCs have been little studied in that respect so far. \cite{Ocko1986, Crawford1996, Kutnjak2003, Binder2008, Mazza2010,Araki2011,Cetinkaya2013, Calus2015, Schlotthauer2015, Ryu2016, Busch2017, Tran2017, Brumby2017, Zhang2019, Sentker2019, Gang2020SoftMatterBook, Monderkamp2021, Salgado2022}  

\iftrue
\begin{figure}[htbp]
 \centering
	\includegraphics[width=1\columnwidth]{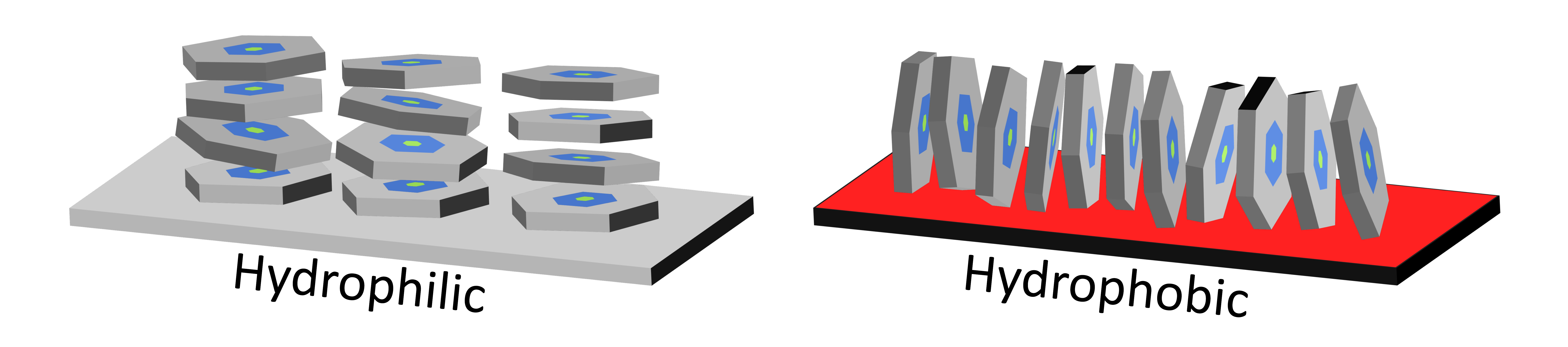}
	\caption{\textbf{Anchoring of discotic ionic liquid crystals at planar surfaces.} (a) face-on (homeotropic) anchoring and (b) edge-on (planar or homogeneous) anchoring of discotic ionic liquid crystal molecules on a hydrophilic and hydrophobic surface, respectively.}
		\label{fig:Anchoring}
\end{figure}
\fi

Several previous studies have been performed on conventional, non-ionic DLCs under nanoconfinement, especially in cylindrical pores of nanoporous anodic aluminum oxide (AAO) templates. \cite{Kopitzke2000, Duran2012, Cerclier2012, Kityk2014, Zhang2014, Zhang2015, Calus2014,Zhang2017, Sentker2018, Yildirim2019, Sentker2019} As a basis for the discussion of our results, we present below the peculiar textures found in these studies. As shown in Fig.~\ref{fig:Anchoring}, due to the hydrophilicity of the aromatic core of the disks and the hydrophobicity at the disk edges, DLC molecules often adopt face-on (homeotropic) anchoring on hydrophilic flat surfaces and edge-on (planar or homogeneous) anchoring on hydrophobic flat surfaces. 

What liquid crystalline textures result from these different anchorings depending on the pore wall hydrophilicities in the cylindrical channels of AAO? For as-prepared AAO with its polar, hydrophilic pore walls, either radial disk arrangements as shown in Fig.~\ref{fig:config}a \cite{Kityk2014} or so-called logpile structures have been deduced, see Fig.~\ref{fig:config}b \cite{Zhang2014, Zhang2015}. The latter are characterized by mostly straight and parallel columns arranged in layers perpendicular to the cylindrical pore axis. Since the stacking orientation can vary along the channel, the texture is reminiscent of logs in an ordered pile. Only near the pore wall do the columns splay slightly. 

\iftrue
\begin{figure}[ht]
 \centering
	\includegraphics[width=1\columnwidth]{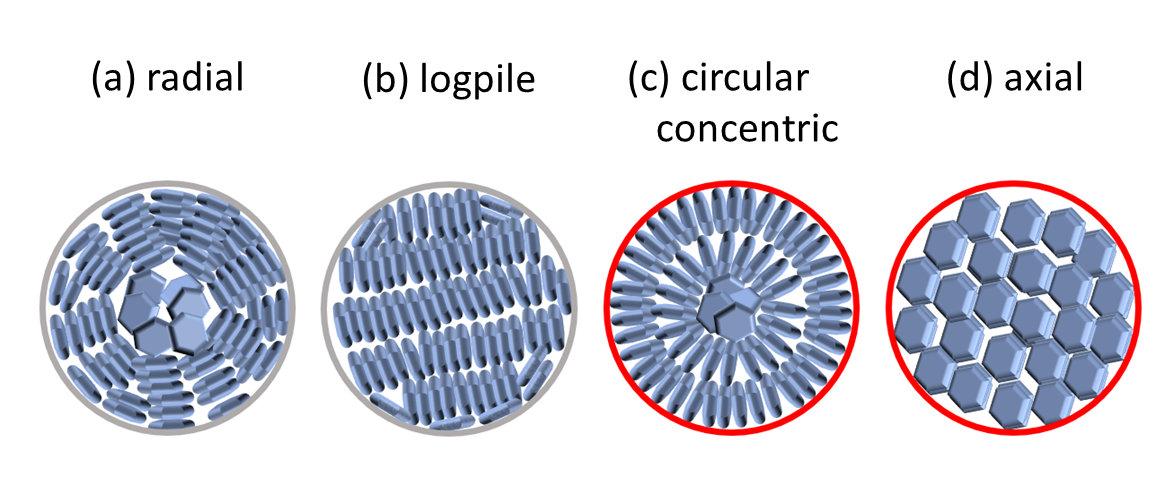}
	\caption{\textbf{Top view of columnar discotic liquid-crystalline order in cylindrical nanopores.} (a) radial and (b) logpile configurations are typical for face-on anchoring at hydrophilic pore walls and (c) circular concentric and (d) axial configurations are typical for edge-on anchoring at hydrophobic pore walls. \cite{Kityk2014, Zhang2014, Sentker2018, Sentker2019}}
		\label{fig:config}
\end{figure}
\fi

In terms of electrical conductivity and thus the suitability for molecular wires, neither configuration is preferred because no percolating conduction paths are established along the cylindrical channel. One might think that hydrophobic anchoring should solve this problem, since an axial alignment of the columns, and thus nicely aligned conduction paths along the long axes of the AAO nanochannels, should be the natural configuration fulfilling the edge-on anchoring condition at the pore wall. However, as first shown by Zhang and Ungar et al\cite{Zhang2015}. in reciprocal space mapping experiments and atomic force microscopy, DLCs often form circular concentric rings instead of straight columns aligned parallel to the long cylindrical AAO pore axes. 

The main cause of these two peculiar liquid crystalline textures, ''circular concentric'' and ''logpile'' and thus the avoidance of an axial arrangement in cylindrical nanoconfinement, be it hydrophilic or hydrophobic, is the very high splay energy and low bending energy of the columnar phases \cite{Zhang2015}. In fact, it could be shown that the axial configuration can be induced by increasing the column bending energy, e.g. by chosing DLCs with large aromatic cores and thus increased column stiffness\cite{Zhang2015}. Another way of achieving the preferred axial arrangement is the selection of small pore sizes, since the elastic cost for the deformation of the hexagons at the pore walls which disfavors this arrangement compared to the circular concentric state, decreases with pore size, so that eventually also for hydrophobic channels an axial arrangement prevails, even for DLCs with relatively small aromatic cores, such as HAT6 \cite{Zhang2015, Sentker2018, Sentker2019, Zhang2017}. 

To what extent these peculiarities with regard to the formation of axially arranged columnar phases are also typical for DILCs confined in AAO is the main focus of the present study. In fact, some previous work has also been done on confined DILCs. Kolmangadi et al. used a combination of dielectric spectroscopy, calorimetry and X-ray scattering to show that the isotropic-to-columnar discotic transition of a DILC appears to be continuous, despite a discontinuous first order bulk phase transition. \cite{Kolmangadi2023Confinement-SuppressedApplications}. However, a detailed understanding of the structure and texture formation of DILCs in confinement is still lacking. 

Here, we investigate the temperature-dependent translational and collective orientational order of DILCs with varying alkyl side-chain length and thus varying hydrophilicity confined in cylindrical nanochannels of 180~nm diameter with different mesogen pore-wall anchoring (hydrophilic, face-on or hydrophobic, edge-on) by synchrotron-based X-ray scattering and high-resolution optical birefringence measurements. Moreover, the DILC have a cyclic or acyclic side group, respectively, where the cyclic group, in contrast to the acyclic one, exhibits a suppressed rotational motion. Thus, the acyclic one shows more rotational dynamics and could thus contribute more disordering effects compared to the cyclic one \cite{Butschies2010,Kolmangadi2021}.

\newpage
\section{Results and discussion}

\subsection{Translational and collective orientational order of cyclic Dopa-ILCs confined in 180~nm AAO membranes}

\iftrue
\begin{figure*}[htbp]
    \centering
    \includegraphics[width=0.73\textwidth]{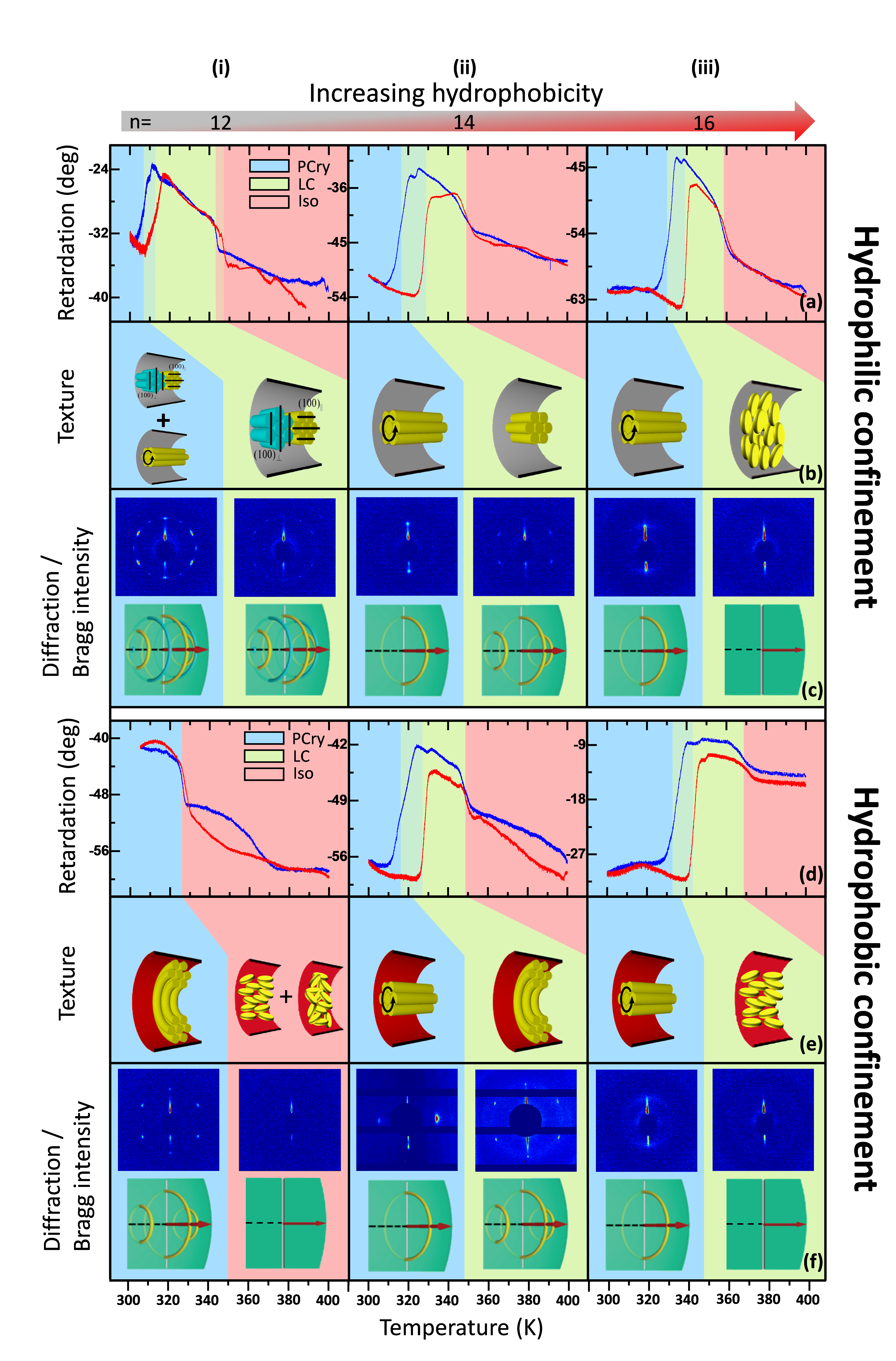}
    \caption{\textbf{Self-assembly of cyclic DILCs as a function of  mesogen hydrophobicity in hydrophilic and hydrophobic nanochannels.} Optical retardation and SAXS patterns of cyclic Dopa-ILC molecules with different side chain lengths $n$ = 12 (i), 14 (ii), 16 (iii) confined in hydrophilic and hydrophobic nanopores with 180~nm pore size. The measured SAXS patterns and schematics of the corresponding scattering geometry and Bragg intensities in reciprocal space intersecting the Ewald sphere in both the plastic crystalline (PCry) and liquid crystalline (LC) phases are given. The background color of the birefringence curves refers to different phases of the DILCs under confinement (blue: PCry phase, green: LC phase, red: isotropic (Iso) phase). The dashed areas at the phase transition are the hysteresis between cooling and heating.}
    \label{fig:cyclicsum}
\end{figure*}  
\fi	

In Fig.~\ref{fig:cyclicsum} we show the birefringence measurements and the small-angle X-ray scattering (SAXS) results of cyclic Dopa-ILCs confined in both hydrophilic (a, b, c) and hydrophobic (d, e, f) nanoporous AAO membranes with a pore size of 180~nm. 

We will start our discussion on the cyclic Dopa-ILC with the shortest side chain length $n$ = 12 (Cy12). As shown in the Fig.~\ref{fig:cyclicsum}a(i), the optical retardation increases as the sample cools down with one relatively sharp increase at about 342~K. This indicates the formation of a collective orientational order where the superdiscs are aligned with their normal perpendicular to the long pore axis. Upon further cooling below 310~K, however, a decrease in the optical retardation indicates a full or at least partial reorientation of the superdiscs' normals parallel to the long cylindrical pore axes. Upon heating this peculiar re-orientational behaviour is inverted and again two orientational ordering transitions are observed with small temperature hystereses of about 5~K between cooling and heating.
        
The X-ray diffraction experiment gives additional important information on the translational order of the superdiscs. Below 342~K we observe a set of 12 diffraction peaks at a modulus of the wave vector transfer $q$ typical of the (100) Bragg peak of hexagonal columnar order. It can be explained by the formation of two distinct domains of logpile structures where the columns grow in radial direction, where however for the one domain the (100) lattice planes are perpendicular to the long pore axis, whereas they are parallel to the long pore axis for the second domain. Along with the rotation symmetry about the long cylindrical axis this results in a set of green and yellow Bragg ring intensities in reciprocal space shown in Fig.~\ref{fig:cyclicsum}c(i). For our geometry and an incident beam perpendicular to the long channel axes, these rings intersect at 12 points with the Ewald sphere resulting in the 12-fold diffraction pattern observed. 

\iftrue
\begin{figure*}[tbp]
	\centering
	\includegraphics[width=1\textwidth]{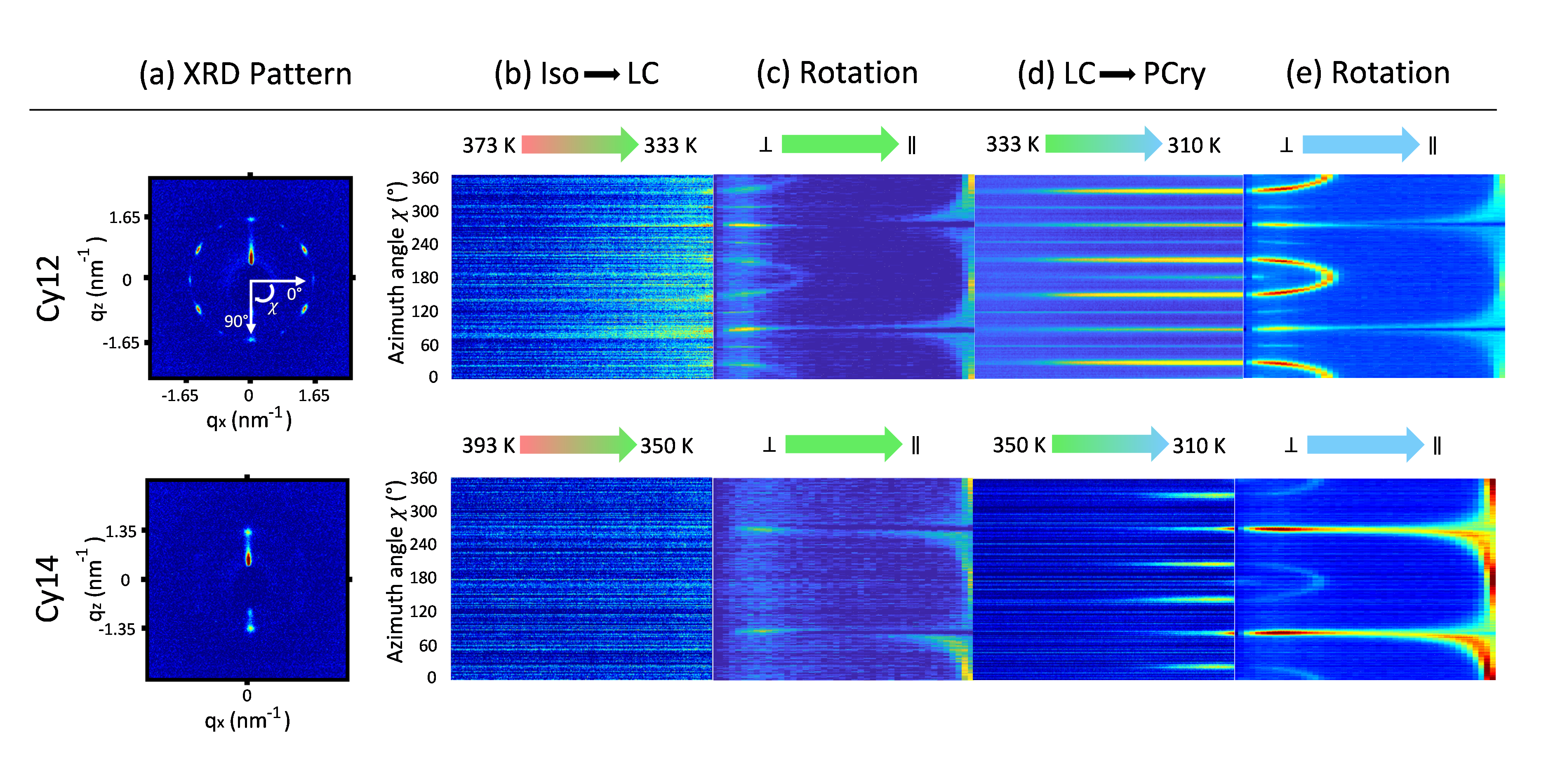}		
	\caption{\textbf{Texture evolution for Cy12 and Cy14 confined in hydrophilic nanochannels upon cooling to the LC and PCry phase.} The upper and lower row contain X-ray scattering data with regard to Cy12 and Cy14, respectively. (a) X-ray diffraction patterns at T= 310~K in the PCry phase. The peaks indicate the wave vector transfer $q_{\rm (10)}$ characteristic of the hexagonal intercolumnar order. (b) Plotted is the color coded scattering intensity at a fixed modulus of the wave vector transfer $q_{\rm (10)}$ as a function of azimuth angle $\chi$ (vertical) for a fixed $\omega = 85^\circ$ upon cooling from $T_{Iso\rightarrow LC}$, (c) in the LC phase upon rotation of the membrane from a perpendicular to a parallel orientation of the pore axis $\hat{p}$ with regard to the beam and (d, e) analogous scans upon cooling from $T_{LC\rightarrow PCry}$ and sample rotation in the PCry phase.}
 
	\label{fig:Azi}
\end{figure*}
\fi	

Fig.~\ref{fig:Azi} plots the Bragg peak intensity of Cy12 and Cy14 confined in 180~nm hydrophilic pores at each azimuth angle during cooling and rotational scans. When the samples are in the high temperature Iso phase, their XRD patterns show no diffraction peaks from ILC structures, only membrane reflections at $\chi = 90^\circ$ and $270^\circ$. To eliminate the membrane reflections, the membrane reflection SAXS peaks at Iso phase are subtracted from all other temperatures. In Fig.~\ref{fig:Azi}(b), at the LC phase temperature a rotational scan is performed where we can see a very weak 12-fold pattern in Cy12 and a pair of equatorial peaks in Cy14 as discussed above. Upon further cooling into the PCry phase, the 12-fold pattern of Cy12 becomes stronger both in the cooling scan of $T_{LC\rightarrow PCry}$ and in the rotation scan in the PCry phase. Interestingly, from the Bragg intensity distribution it can be seen that the peaks resulting from the (100)$\parallel$ domains ($\chi = 30^\circ,$ $90^\circ,$ $ 150^\circ,$ $ 210^\circ,$ $ 270^\circ$ and $330^\circ$) are more prominent than the remaining peaks resulting from the (100)$\perp$ domains, indicating that the (100)$\parallel$ domains dominate in the pores over the (100)$\perp$ domains. On the other hand, upon further cooling to the Cry phase, the equatorial peaks of Cy14 changes to a hexagonal pattern at around 325~K, indicating the structural transformation from an axial structure to a logpile structure.
        
Thus, both our optical and X-ray experiment consistently indicate the formation of logpile structures upon cooling from the isotropic phase. Upon cooling below the second transition in the optical experiments we see barely any change in the diffraction patterns, except for an increase in the intensity of the (100)$\parallel$ domain. By contrast, the decrease in optical retardation hints at least to a partial axial orientation of the columns. Therefore, we infer a coexistence of axial and logpile columnar textures, see Fig.~\ref{fig:cyclicsum}b(i) and c(i). 
        
These findings are different from the observations for the more hydrophobic mesogens Cy14 and Cy16. For both systems, analogous to Cy12, we see first an increase in optical retardation and then a sharp drop. However, two Bragg peaks in equatorial directions (at $\chi=$ 90$^\circ$ and 270$^\circ$) at positions typical of the intercolumnar distance are dominant at low $T$ in the PCry phase. These equatorial peaks at low $T$ can be explained by one ring in reciprocal space which results from a hexagonal columnar order along the long pore axis with randomization of the hexagonal orientation about the pore axis, see Fig.~\ref{fig:cyclicsum}b(ii) and c(ii). Thus we have evidences of axial order at low temperature in the PCry phase for both mesogens. 
In the lower panels of Fig.~\ref{fig:Azi} these findings are documented in more detail by temperature- and rotation- dependent contour plots of the $q_{\rm (10)}$ intensity for Cy14. We observe in the intermediate $T$-range (between the LC and the PCry phase, 350 K to 330 K) only a six-fold pattern typical of the (100)$\parallel$ domain, no coexistence with (100)$\perp$ domains. Note that these scans reveal that additionally to the dominating axial peaks also scattering intensity emerges at positions typical of the logpile structure, see panel (d) and (e). Thus, also a presumably very small fraction of logpile textures are present in coexistence with the dominating axial arrangment.

Interestingly, for the Cy16 mesogen no translational order at all can be inferred in the LC phase, despite the increase in collective radial orientational order indicated by the polarimetry experiments, see Fig.~\ref{fig:cyclicsum}c(iii). Hence, the system forms a discotic nematic phase. 
        
Overall we find quite some analogies with classical discotic systems in terms of the formation of logpile structures with preferred hexagonal axis orientation with regard to the long pore axis. Moreover, by increasing the hydrophobicity (chain length) of the mesogens the systems show an increasing tendency to perform a textural transition from a logpile configuration to axial order. These textural transitions are consistent with LC to PCry transitions of the bulk and are therefore likely to be triggered by improved columnar rigidity. Presumably, this is caused by a gradual freezing (increasing all-trans configuration) of the alkyl-side chains upon cooling. Such freezing effects of the alkyl side chains can be observed in many discotic liquid crystal systems when cooled from the LC phase to the crystalline phase. In this study, on slow cooling to the PCry phase, the three alkyl chains of Dopa-ILCs gradually lose their flexibility as the chain stiffness increases, resulting in an increasing intercolumnar spacing. We exemplify this behaviour in Fig.~\ref{fig:negative_expansion} for Cy12 in hydrophilic pores. Note that the overall improved hexagonal columnar order is also evidenced by the increase in the $q_{\rm (10)}$ peak intensities in the PCry phase.

\iftrue
\begin{figure}[htbp]
 \centering
	\includegraphics[width=0.8\columnwidth]{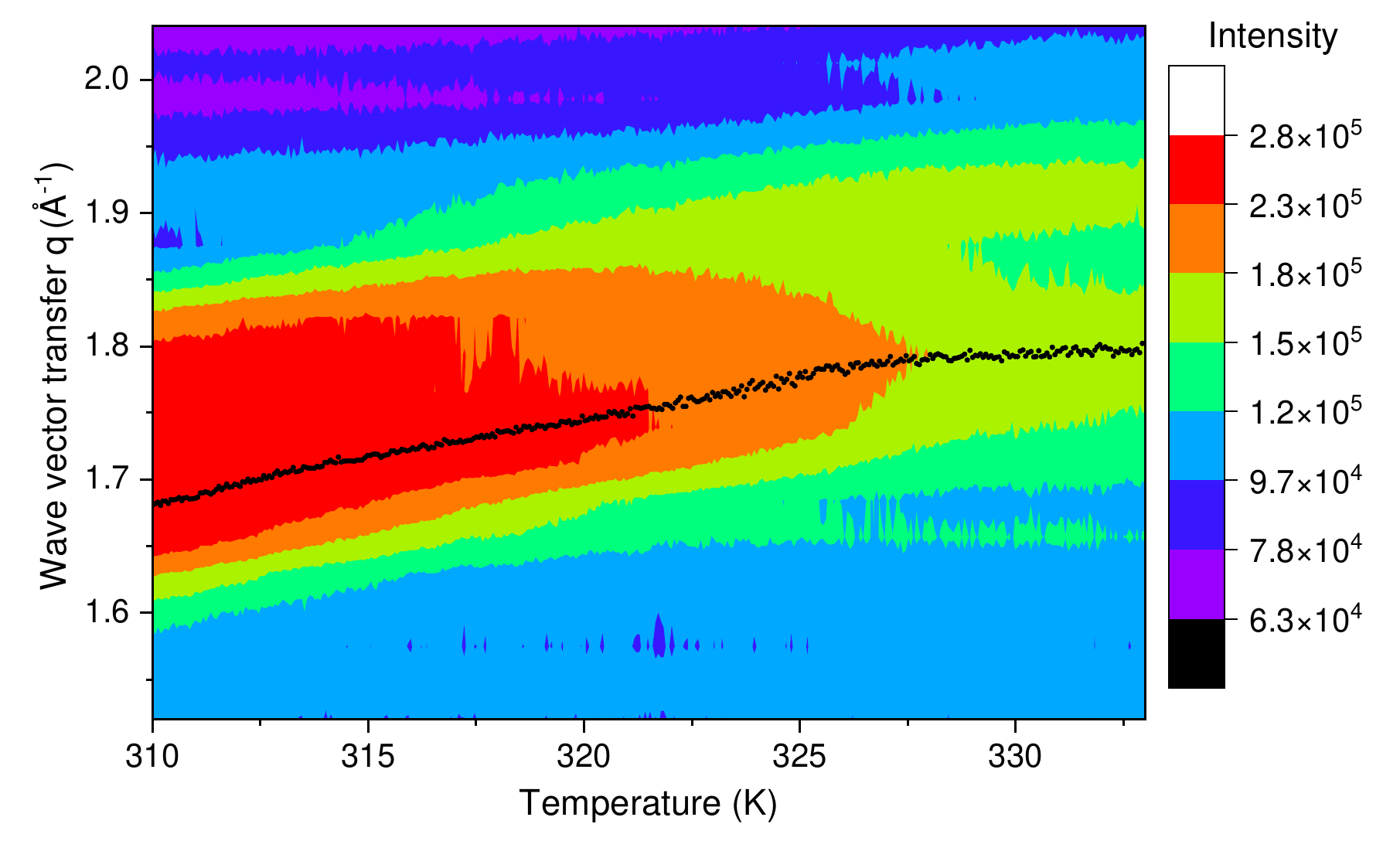}
	\caption{\textbf{Negative thermal expansion of the intercolumnar lattice.} Plotted is the color-coded X-ray scattering intensity as a function of temperature in the vicinity of the wave vector transfer of the dominating hexagonal (10) Bragg peak position $q_{\rm (10)}=4\pi\,/(\sqrt{3}d$), where $d$ is the intercolumnar lattice constant $d$, as a function of temperature. As a guide for the eye the position of the maximum intensity is indicated as black line. It increases with increasing temperature indicating a Bragg peak shift to higher $q$ with increasing temperature and thus an (unconventional) decreasing lattice parameter (negative thermal expansion).}
		\label{fig:negative_expansion}
\end{figure}
\fi

Particularly noteworthy is also the observation of a confinement-induced nematic discotic phase in the case of Cy16, since such a phase has been barely found in other DILC systems in the bulk state. \cite{Kapernaum2022} 

For the hydrophobic confinement two orientational transitions can be inferred from the optical birefringence experiments for each mesogen, see~Fig.~\ref{fig:cyclicsum}. However, for the most hydrophilic one (Cy12) no decrease in the optical retardation at lower temperature is observed, but rather an additional increase and thus increased radial orientation of the superdisc normals at low temperature. By contrast for the more hydrophobic molecules both the diffraction experiments (observation of two Bragg peaks) and decrease in optical retardation indicate the formation of axial oriented columns at low temperature. 
        
It is also interesting to mention that except for Cy14 no diffraction patterns (Bragg peaks) can be detected upon the first optical retardation increase. This means that in both cases only collective orientational order but no translational columnar order is emerging upon cooling from the isotropic state. The systems form nematic discotic states. 

Note that we indicate for Cy12 at low and Cy14 for intermediate temperatures a circular concentric columnar ring formation, in analogy to the analogous observation for classic discotic systems, see Fig.~\ref{fig:cyclicsum} e(i)\&(ii). This conclusion is drawn from the observation of a six-fold pattern in combination with the boundary condition of an edge-on anchoring of the discs at the pore walls, see Fig.~\ref{fig:cyclicsum} f(i)\&(ii). Unfortunately, both the circular concentric ring formation as well as the logpile structures result effectively in the same qualitative Bragg ring distribution in reciprocal space, since both emerge from the rotation of a hexagon about the pore axis in reciprocal space. \cite{Zhang2014, Sentker2019} Whereas for discotic liquid crystal AFM studies could confirm this structure formation for edge-on anchoring, we can only indirectly infer this. \cite{Zhang2014} Also a reasoning with regards to correlation lengths did not provide a more conclusive result. 
        
In summary, the cyclic Dopa-ILCs confined in 180~nm hydrophilic and hydrophobic pores show a clear dependence on the molecular hydrophobicity (side chain length). ILCs with shorter side chains and thus a dominance of the hydrophilic core tend to orient radially with uniform collective orientation and translational order in both the LC and PCry phases under confinement, whereas larger molecules with longer side chains tend to orient radially only in their LC phase. As the temperature decreases into the PCry phase, the larger molecules adopt an axial configuration instead to avoid the high bending energy and spatial distortion due to the freezing of their long side chains. This emphasizes again the importance of the bending elasticity for the texture formation in nanoconfinement.

\subsection{Translational and collective orientational order of acyclic Dopa-ILCs confined in 180~nm AAO membranes}

\iftrue
\begin{figure*}[htbp]
    \centering
   \includegraphics[width=0.73\textwidth]{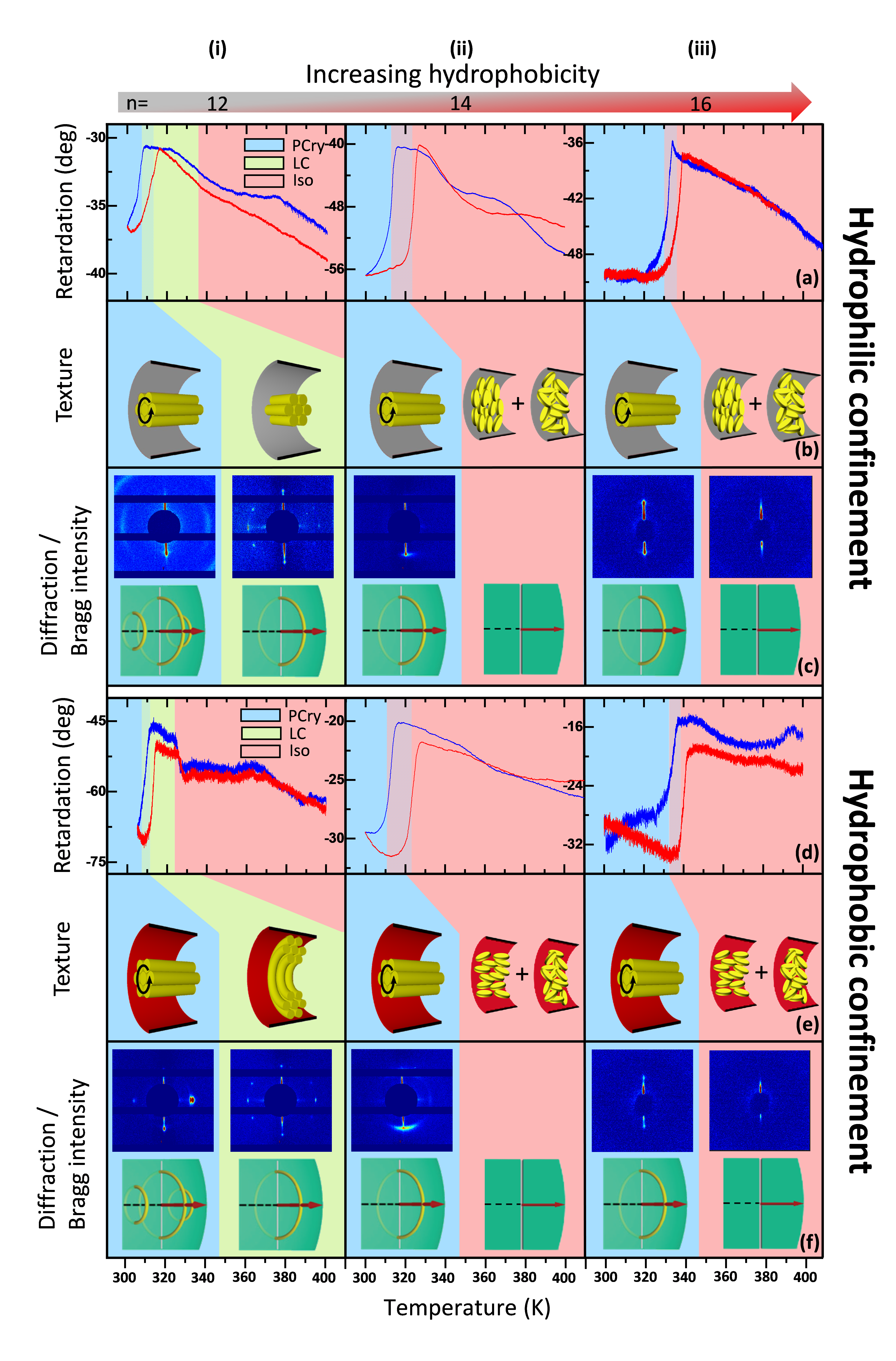} 
    \caption{\textbf{Self-assembly of acyclic DILCs as a function of mesogen hydrophilicity in hydrophilic and hydrophobic nanochannels.} Birefringence curves and XRD patterns of acyclic DOPA-ILC molecules with different side chain lengths confined in hydrophilic and hydrophobic nanopores with 180~nm pore size at different phases. The measured SAXS patterns and schematics of the corresponding scattering geometry and Bragg intensities in reciprocal space intersecting the Ewald sphere in both the plastic crystalline and liquid crystalline phases are also given. The background color of the birefringence curves refers to different phases of the ILCs (blue: PCry phase, green: LC phase, red: Iso phase). The dashed areas at the phase transition refer to the hysteresis between the cooling and heating processes.}
    \label{fig:Acyclicsum}
\end{figure*} 
\fi

In Fig.~\ref{fig:Acyclicsum} we show the SAXS measurement of the acyclic Dopa-ILCs confined in both hydrophilic and hydrophobic AAO membranes with 180~nm pore sizes. For the acyclic Dopa-ILC with the shortest side chain length $n$ = 12 (Ac12) confined in both hydrophilic and hydrophobic pores as shown in Fig.~\ref{fig:Acyclicsum}(i), six-fold diffraction patterns are first observed in their LC phase. This indicates a logpile structure in hydrophilic pores and a circular concentric structure in hydrophobic pores as already discussed in the previous section. If the sample is further cooled to the PCry phase, the six-fold pattern evolves into a pair of equatorial peaks, which is the typical diffraction pattern for an axial structure. Therefore, when the Ac12 sample is cooled from the LC to the PCry phase, an interesting structural transition from radial to axial structure is again observed. A similar conclusion can also be drawn from the birefringence curves as shown in Fig.~\ref{fig:Acyclicsum}a(i)\&d(i). For both hydrophilic and hydrophobic membranes, the retardation value first increases on cooling from the isotropic phase to the LC phase, forming a logpile/circular concentric structure in the nanopores. Then, similarly as for the cyclic DILCs on further cooling to the PCry phase, the side chains of the ILC molecules freeze, resulting in higher chain stiffness. An axial structure is energetically preferred to avoid the high bending and splay distortion of the logpile/circular concentric configuration. As a result, when cooled from the LC phase to the PCry phase, Ac12 shows a decrease in the retardation value caused by the radial to axial configuration transformation. 
        
For the DILCs with acyclic functional groups the molecular hydrophobicity increases with increasing alkyl side chain lengths. When confined in both hydrophilic and hydrophobic pores as shown in Fig.~\ref{fig:Acyclicsum}(ii) and (iii), both Ac14 and Ac16 first show no obvious diffraction pattern in the LC phase and then the polar peaks appear in the PCry phase, indicating no molecular translational order in the LC phase and an axial configuration in the PCry phase. The same conclusion can also be drawn from the birefringence curves, where the retardation values show no obvious boundaries between the isotropic phase and LC phase and a sharp decrease when cooled down to the PCry phase. 

\subsection{Landau-de Gennes analysis of the isotropic-to-liquid-crystal transitions}

\iftrue
\begin{figure*}[htbp]
    \centering
   \includegraphics[width=0.75\textwidth]{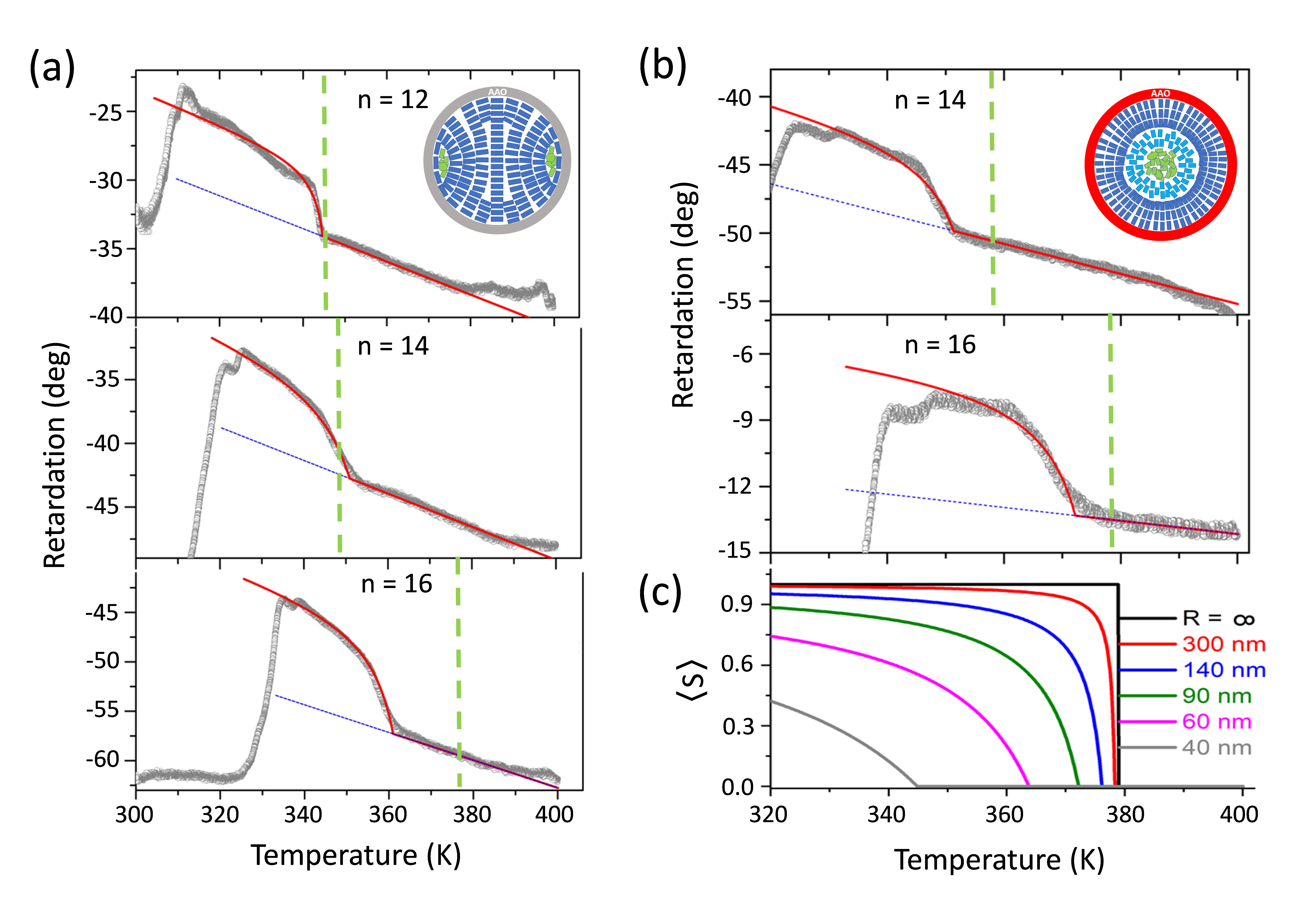} 
    \caption{\textbf{Landau-de-Gennes modelling of the cyclic Dopa-ILCs phase behaviour in (a) hydrophilic and (b) hydrophobic nanopores.} The green dashed line shows the first-order bulk phase transition temperature and the red curve shows the fitting curve using the Landau-de-Gennes theory. Inset: (a) logpile configuration of Dopa-ILCs in hydrophilic pores and (b) circular concentric configuration of Dopa-ILCs in hydrophobic pores. (c) Simulations on the dependence of effective orientational order parameter $\langle S \rangle$ of Dopa-ILCs confined in different pore sizes.}
    \label{fig:Fitting_Landau}
\end{figure*} 
\fi

The optical retardation is a direct measure of the liquid crystalline order parameter $\langle S \rangle$. Whereas for the acyclic systems barely any systematic thermotropic changes in the optical birefringence between the Iso and LC phases can be detected, for the cyclic systems quite pronounced changes in the retardation are visible. In all cases, it shows a continuous transition behaviour in the region of the Iso to LC phase transition $T_{Iso\rightarrow LC}$, which contrasts with the discontinuous jump-like behaviour of the order parameter characteristic of the first-order transition observed in the corresponding bulk systems, as seen for example the evolution of the birefringence of a bulk film of Cy12, Cy14 and Cy16 in the appendix. This  observation is also in agreement with the findings in our previous dielectric and calorimetric study. \cite{Kolmangadi2023Confinement-SuppressedApplications}

In general, such a behaviour can be rationalized by considering two mechanisms: (i) interfacial anchoring and (ii) inhomogeneous deformations of the director fields caused by cylindrical confinement, which dominate in different temperature regions, i.e. in the confined Iso and LC phases, respectively. 

The molecular ordering due to the Iso to LC transition contributes to the excess birefringence and is linearly related to its optical retardation. It appears on the background of the temperature-dependent birefringence. As a reasonable approximation, the bare geometrical retardation is assumed to be linearly dependent on temperature, so in most analyses its value is extrapolated from the parent phase, in this case the isotropic one. The excess retardation can be positive or negative depending on the optical molecular anisotropy and molecular arrangement. Our X-ray scattering experiments suggest a logpile face-on configuration for hydrophilic cylindrical channels and the circular concentric configuration for hydrophobic channels. Therefore, the excess optical retardation is expected to be positive in both cases, which is consistent with the polarimetry measurements.

        %Hydrphilic

Homeotropic (or face-on) anchoring of discotic Dopa-ILC building blocks to the hydrophilic cylindrical walls of the host matrix occurs well above $T_{Iso\rightarrow LC}$. As the temperature decreases the anchoring effects strengthen as a result of the increase in the number of statistically anchored molecules at the pore walls in the confined isotropic phase. For rod-like nematic systems this typically results in a pretransitional tail and in a paranematic state. It is associated with a geometrical ordering field, $\sigma \propto 1/R$, enforcing arrangement of rod-like molecules along the channel long axis \cite{KKLZ1,KKLZ2}. 
Under increasing confinement such nematogen systems reveal a crossover from discontinuous to continuous temperature evolution of their orientational (nematic) order parameter with a characteristic critical point \cite{KKLZ1,KKLZ2,SC1,SC2}. This mechanism, however, is not fully applicable to discotic molecules homeotropically anchored on cylindrical pore walls: The optical polarizability of discotic molecules is isotropic in their aromatic plane thus any lateral rotation of them keeps the effective optical anisotropy of the LC composite unchanged despite the fact that face-on bonding indeed takes place giving certain contribution to the optical retardation in the paranematic phase. Accordingly, there are no pre-transitional tails observable. The linearly with temperature increasing birefringence above the transition is traceable to the purely linear increase in geometric birefringence, see blue lines in Fig. \ref{fig:Fitting_Landau}.
        
The most probable reason for a continuous evolution of the effective orientational order parameter $\langle S \rangle$, likewise related to it retardation ($R \propto \langle S \rangle$) is the inhomogeneous bend and splay deformation of the columns caused by cylindrical confinement. For face-on anchoring with logpile arrangement in the pore centre this is schematically shown in the inset of Fig.~\ref{fig:Fitting_Landau}(a). Note that we emphasize in this drawing the distinct kinds of deformations for clarity. Besides that, in the small area between the columns and the pore wall there also exist some randomly distributed molecules. The splay/bend deformation as well as the small random area all introduce excess energy and contribute to the shift of the phase transition temperature. Unfortunately, the resulting inhomogeneous director field along with the unknown knowledge of the detailed relative contributions of splay and bend deformation along the nanochannel axis and in the cross section hamper a sound analytical treatment of the defect energy contributions. Interestingly, however, a fitting of the continuous behaviour by a Landau-de Gennes excess free energy that solely considers the splay deformations at the walls already results in a very good agreement with the continuous evolution of the order parameter, see the fits in  Fig.~\ref{fig:Fitting_Landau}(a). A detailed description of the corresponding model is given in chapter III.E of Ref. \cite{Kityk2014}. 

        %Hydrophobic
Fortunately, edge-on anchoring, which is characteristic of hydrophobically grafted cylindrical AAO channel walls, leads to a theoretically more tractable director field in the case of circular concentric column formation. The formation of circular columns results in bending stresses, $ |\vec{n} \times (\vec{\nabla}\times \vec{n})| = 1/r$, while the splay and twist distortions are both zero \cite{Zhang2014, Sentker2018}. The biquadratic coupling between the orientational order parameter $\langle S \rangle$ and the bending distortion [$b_3(\vec{n} \times (\vec{\nabla}\times \vec{n}))^2S^2$-term] causes a distinct, local transition temperature $T_c(r) = T'(R) - b_3A^{-1}r^{-2}$ for each circular concentric ring with ring radius $r$. According to this formalism, circular concentric layers start to form near the channel walls during cooling at $T_{ICD} = T'(R) - b_3A^{-1}r^{-2}= T^{\#}_{ICD} - g/R-b_3A^{-1}R^{-2}$. The radius of the cylindrical phase front $r_c$ separating the isotropic core from the circular columnar shell is defined as $r_c(T) = [b_3A^{-1}/(T'(R)-T)]^{1/2}$, i.e. it gradually shrinks towards the channel centre as it cools. However, it can be seen that even at $T \rightarrow 0$, $r_c(T=0) = [b_3A^{-1}/T'(R)]^{1/2}$: the radius of the isotropic core tends to a finite non-zero value. Therefore, it can be assumed that the circular concentric arrangement of discotic columns is characterized by gradual frustration as one moves from the periphery to the centre of the channel, first losing columnar (positional) order, resulting in the formation of an intermediate bent nematic layer, and finally losing orientational order as one approaches the isotropic core, see sketch in inset of Fig.~\ref{fig:Fitting_Landau}(b). Assuming fully saturated order in the subsequent circular columns and in the bent nematic layer, the effective order parameter in the relevant phase is defined as $\langle S(T)\rangle = 1-b_3A^{-1}R^{-2}(T'(R) - T)^{-1}$. Fig.~\ref{fig:Fitting_Landau}(b) shows the temperature dependence of the optical retardation $\Delta(T)$ (grey dots) measured on cooling (hydrophobic grafting, see Fig.~\ref{fig:Acyclicsum}(d) and corresponding fits (red curves) for comparison. And  Fig.~\ref{fig:Fitting_Landau}(c) shows simulated $\langle S(T)\rangle$ dependencies ($R=\infty$, 300, 140, 90, 60 and 40 nm). Such a bend distortion model reproduces the temperature dependence of the optical retardation close to the phase transition point well, in particular its rounded kink-like character. Note that this reasoning is also consistent with the conclusions regarding a DLC (HAT6) confined in hydrophobic silica nanopores. \cite{Sentker2018} Due to the large ratio of molecule size to pore diameter, the bending energy differences of the circular ring layers with distinct radii are even so large for this system that a layer-by-layer quantized transition behaviour has been observed both in experiment and in Monte Carlo computer simulations. \cite{Sentker2018}

\section{Conclusions}
We presented temperature-dependent high-resolution optical birefringence measurements and 3D reciprocal space mappings based on synchrotron-based X-ray scattering to investigate the thermotropic phase behaviour of dopamine-based ionic liquid crystals in cylindrical channels of 180~nm diameter in anodic aluminum oxide membranes. As a function of the hydrophilicity and thus the molecular anchoring to the pore walls (edge-on or face-on) and the variation of the hydrophilic-hydrophobic balance between the aromatic cores and the alkyl side chain motifs of the superdiscs, we find a particularly rich phase behaviour, which is not present in the bulk state.

Comparing the cyclic and acyclic molecules with the same side chain length, it can be inferred that the acyclic five-membered ring as a functional group introduces a higher molecular hydrophobicity than the cyclic five-membered ring. For example, Cy14 still shows a radial structure while Ac14 forms only a less ordered paranematic structure in its LC phases.

For both cyclic and acyclic Dopa-ILCs with different side chain lengths, it can be concluded from the birefringence curves and SAXS patterns that DILCs with lower molecular hydrophobicity (shorter side chain) tend to orient radially under cylindrical confinement, whereas DILCs with higher molecular hydrophobicity (longer side chain) tend to form a paranematic structure in the LC phase. This confinement-induced nematization is remarkable, since in the DILC bulk state nematic states are found rather rarely.

Also confinement-induced continuous order formation is observed in contrast to discontinuous first-order phase transitions, which can be quantitatively described by Landau-de Gennes free energy models for liquid crystalline order transitions in confinement.

Upon cooling from the LC phase to the PCry phase, presumably due to the freezing of the alkyl side chains and the consequent increase in molecular stiffness and hydrophobicity, almost all ILCs adopt an axial structure to avoid the high bending/splay energy resulting from the radial structure. Among them, Cy14 and Ac12 show particularly interesting structural behaviour upon cooling from LC to PCry phase. Both give clear indications of the structural transition from logpile (hydrophilic) or circular concentric (hydrophobic) orientation to axial orientation with decreasing temperature. Thus, the DILC system presented here is a fine example, where a cooling-induced increase in bending rigidity of discotic columns drives a textural transition from a radial to an axial self-assembly in nanoconfinement. 

In terms of applications, the temperature- and confinement-dependent radial-to-axial phase transition as well as nematization could be very interesting for the design of nanoporous functional devices and membranes with tunable ionic conductivity along the tubular nanopores or for optical hybrids with tunable birefringence.

\section{Experimental}

\subsection{Materials}

The synthesis of the investigated DILCs is described in Ref. \cite{C8CP03404D}. Their multi-scale self-assembly is illustrated in Fig.~\ref{fig:assem}. The single ionic liquid crystal molecular block is based on L-3,4-dihydroxyphen-ylalanine (Dopa-ILC). It consists of a large organic cation with an aromatic core (blue), three alkyl side chains, as well as a chloride counterion (green). Dopa-ILCs with two different functional groups as cationic units (cyclic or acyclic imidazolium five-membered ring) are investigated and in both cases the chain lengths of the alkyl side chains are varied ($n$ = 12, 14 and 16). A single Dopa-ILC molecule has a conical shape due to the presence of three flexible alkyl chains, resulting in inverted micellar-type self-assembled discs (Fig.~\ref{fig:assem} (b)), where each disc is composed of six Dopa-ILC mesogens. \cite{C8CP03404D} Given the chemical structure of the disc units, each disc should be hydrophilic in the centre and hydrophobic at the edge. Driven by the $\pi-\pi$ interactions between the aromatic cores, the discs can further self-organise and stack in columns, leading to a hexagonally ordered columnar liquid crystalline mesophase (Fig.~\ref{fig:assem} (c) and (d)). \cite{Yildirim2018}

Nanoporous AAO membranes purchased from SmartMembranes GmbH (Halle, Germany) were used as the host material to confine the Dopa-ILCs. The membranes are 1~cm~$\times$~1~cm square pieces with a thickness of 100~$\upmu$m. Cylindrical pores with a pore diameter of 180~nm are hexagonally distributed on the membrane. The native AAO membranes have hydrophilic pore walls. To investigate the influence of surface chemistry on the self-assembly behaviour, the AAO membranes are also modified by chemical treatment with octadecylphosphonic acid (ODPA). \cite{Grigoriadis2011} During the ODPA treatment, hydrophilic O-H surface groups are replaced by hydrophobic $\mathrm{P-{O_3}-{({CH_3})_{18}}}$ ones, resulting in hydrophobic pore surfaces.

\subsection{Sample preparation} 

DILCs are confined into nanoporous AAO membranes via spontaneous imbibition. \cite{Gruener2011} The membranes are first degassed at 200\,$^\circ$C for 20~h. They are then placed on top of DILCs at a temperature of 20~K above the bulk columnar-to-isotropic phase transition point of the corresponding liquid crystal for 48~h in an argon environment. The Laplace pressure causes a spontaneous infiltration of the molten liquid crystals into the nanoporous membrane. After filling the pores, the remaining bulk material on the surface of the membranes is carefully removed with a razor blade.

As shown in Fig.~\ref{fig:Anchoring} the native nanoporous AAO membranes are hydrophilic, which favors face-on (homeotropic) anchoring of the discs on the pore walls. After chemical modification with ODPA, the pore walls become hydrophobic, which favors edge-on (planar or homogeneous) anchoring of the discs. Although the chemical modification with ODPA grafts a layer of alkyl chains onto the pore walls, the liquid crystal molecules can push the alkyl chains towards the pore wall after full infiltration and the alkyl chains only form a 2.2~nm thick layer on the pore wall and therefore, the pore sizes of the chemically modified membranes are considered to be unchanged compared to the native membranes. \cite{Sentker2019}

\subsection{Optical birefringence measurement}
Temperature-dependent optical birefringence measurements are performed to determine the collective thermotropic orientation of the discotic Dopa-ILC units in the nanopores. 

The optical birefringence measurement setup using modulation polarimetry is shown in the Appendix. Throughout the measurement, the temperature of the sample is increased and decreased at a controlled rate (0.15~K/min) to observe the thermotropic effect. By measuring the optical retardation between the ordinary beam and the extraordinary beam, we obtain information about the collective orientation change of the molecules during heating and cooling in the nanopores. During cooling into the liquid crystalline phase, an increase in the retardation value indicates that the rotational axis of the discotic units is collectively perpendicular to the pore axis, while a decrease in the retardation value indicates that the rotational axis of the discotic units is collectively parallel to the pore axis. \cite{Kityk2008,Kityk2009}

\subsection{X-ray scattering experiments}

\iftrue
\begin{figure}[htbp!]
	\centering
	\includegraphics[angle=0,width=0.4\textwidth]{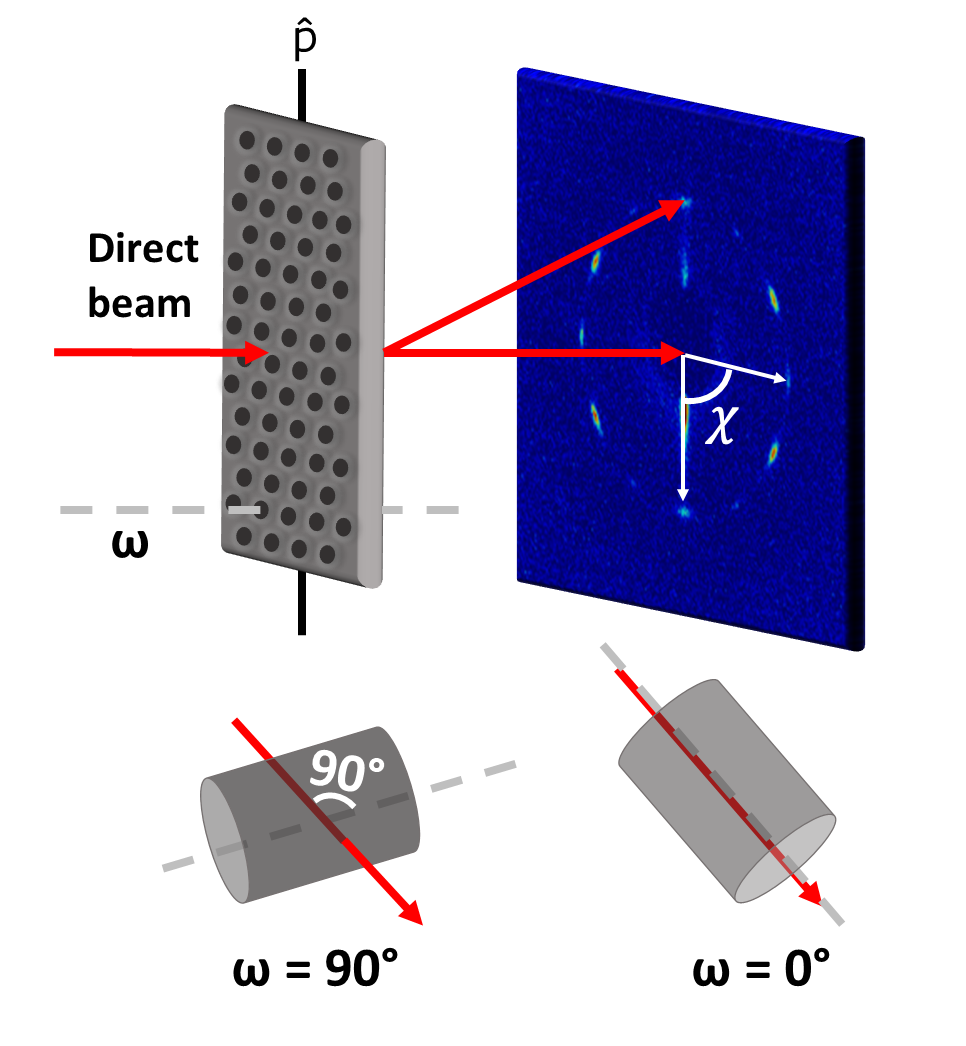}	
	\caption{{\textbf{Schematics of the synchrotron-based X-ray scattering experiment.} Shown are configurations with the long pore axis direction $\hat{p}$ perpendicular and parallel to the incident (direct) beam, which corresponds to sample rotation angles $\omega=90 ^\circ$ and $\omega=0 ^\circ$, respectively. Also indicated is the definition of the azimuth angle $\chi$.}}
	\label{fig:3DRezScattering}
\end{figure}
\fi

To obtain detailed information about the molecular packing of the DILCs, temperature-dependent (1\,K/min) X-ray scattering experiments in transmission were performed at the P08 beamline \cite{Seeck2012} of the PETRA III synchrotron at Deutsches Elektronen-Synchrotron DESY (beam size (VxH) = (200$\times$200)\,$\upmu$m$^2$, beam wavelength $\lambda = $ 0.496\,\AA, Perkin Elmer detector), see Fig.~\ref{fig:3DRezScattering} for a sketch of the scattering geometry and used naming conventions. The temperature-dependent X-ray scattering patterns are taken at $\omega=$85$^\circ$, where the X-ray beam is almost perpendicular to the long axis of the pores,  $\hat{p}$. Additional rotation scans are performed at selected temperatures for sample rotation angles from $\omega$ =  $90^\circ \rightarrow 0^\circ$. For $\omega = $ 85$^\circ$~only order along $\hat{p}$ is probed, while $\omega = $ 0$^\circ$~probes dominantly order perpendicular to $\hat{p}$ since the scattering vector $\vec{q}$ is nearly perpendicular to the beam direction at small scattering angles. Similar to fibre diffraction, we define the $\chi = 0^\circ$~axis as the polar direction and the $\chi = 90^\circ$~axis as the equatorial direction since we consider the long pore axis as equivalent to the fibre direction \cite{Zhang2014, Sentker2019}. By performing Gaussian lineshape fits to the scattering profiles the full width at half maximum (FWHM) is extracted and in turn enables a calculation of the average domain sizes through the Scherrer equation. Here, the domains size is described as $\chi=K\cdot\lambda/\Delta \theta\cdot\cos(\theta)$ where $\lambda$ is the wavelength, $\theta$ is the Bragg angle, $\Delta \theta$ is the fit obtained FWHM and $ K = 0.9$ is an empirical proportionality factor. Using this approach both the coherence lengths along the pore axis ($\chi_\parallel$) and perpendicular to the pore axis ($\chi_\perp$) are determined.

\begin{acknowledgement}

Funding by the Deutsche Forschungsgemeinschaft (DFG, German Research Foundation) within the project ''Ionic Liquid Crystals Confined in Nanoporous Solids: Self-Assembly, Molecular Mobility and Electro-Optical Functionalities'', Projektnummer 430146019 as well as the collaborative research centre CRC 986 ''Tailor-Made Multi-Scale Materials Systems'', Projektnummer 192346071 is acknowledged. We thank Deutsche Elektronen-Synchrotron DESY, Hamburg for access to the beamline P08 of the PETRA III synchrotron and the European Synchrotron Radiation Facility (ESRF) for beamtime at BM02. The presented results are part of a project that has received funding from the European Union's Horizon Europe research and innovation programme under the Marie Skłodowska-Curie Grant agreement no. 101086493. A.V.K. acknowledges the project co-financed by the Polish Ministry of Education and Science under the program "Co-financed international projects", project no. W26/HE/2023 (Dec. MEN 5451/HE/2023/2).

\end{acknowledgement}

\begin{suppinfo}

Experimental procedures and characterization data for all new compounds are provided in the supporting information.

\end{suppinfo}

%\bibliography{ConfinedILC_Li}
\providecommand{\latin}[1]{#1}
\makeatletter
\providecommand{\doi}
  {\begingroup\let\do\@makeother\dospecials
  \catcode`\{=1 \catcode`\}=2 \doi@aux}
\providecommand{\doi@aux}[1]{\endgroup\texttt{#1}}
\makeatother
\providecommand*\mcitethebibliography{\thebibliography}
\csname @ifundefined\endcsname{endmcitethebibliography}
  {\let\endmcitethebibliography\endthebibliography}{}

\end{document}